\documentclass[twoside,twocolumn,9pt]{article}

\usepackage[utf8]{inputenc}
\usepackage{subfig}
\usepackage{t1enc,latexsym,amsfonts,amssymb,amsmath}
\usepackage{bm}
\usepackage{graphics}
\usepackage{wasysym}
\usepackage{setspace}
\usepackage{url}

\usepackage{extsizes}
\usepackage[super,sort&compress,comma]{natbib}
\usepackage[version=3]{mhchem}
\usepackage[left=1.5cm, right=1.5cm, top=1.785cm, bottom=2.0cm]{geometry}
\usepackage{balance}
\usepackage{times,mathptmx}
\usepackage{sectsty}
\usepackage{graphicx}
\usepackage{lastpage}
\usepackage[format=plain,justification=justified,singlelinecheck=false,font={stretch=1.125,small,sf},labelfont=bf,labelsep=space]{caption}
\usepackage{float}
\usepackage{fancyhdr}
\usepackage{fnpos}
\usepackage[english]{babel}
\usepackage{array}
\usepackage{droidsans}
\usepackage{charter}
\usepackage[T1]{fontenc}
\usepackage[usenames,dvipsnames]{xcolor}
\usepackage{setspace}
\usepackage[compact]{titlesec}
\usepackage{color}

\usepackage{blindtext}

\usepackage{epstopdf}

\usepackage{dcolumn}   

\definecolor{cream}{RGB}{222,217,201}

\begin{document}

\pagestyle{fancy}
\thispagestyle{plain}
\fancypagestyle{plain}{

\fancyhead[C]{\includegraphics[width=18.5cm]{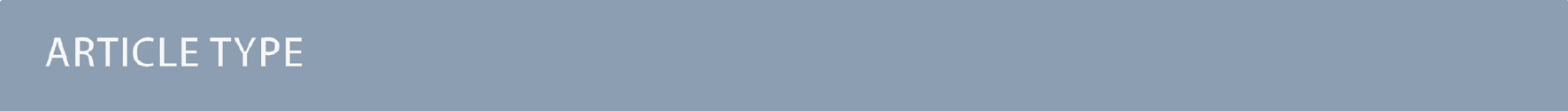}}
\fancyhead[L]{\hspace{0cm}\vspace{1.5cm}\includegraphics[height=30pt]{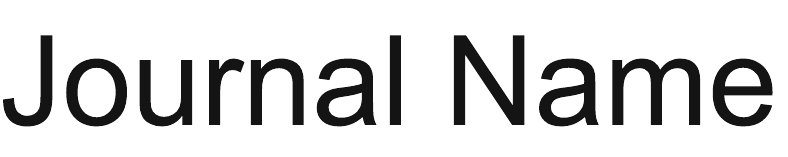}}
\fancyhead[R]{\hspace{0cm}\vspace{1.7cm}\includegraphics[height=55pt]{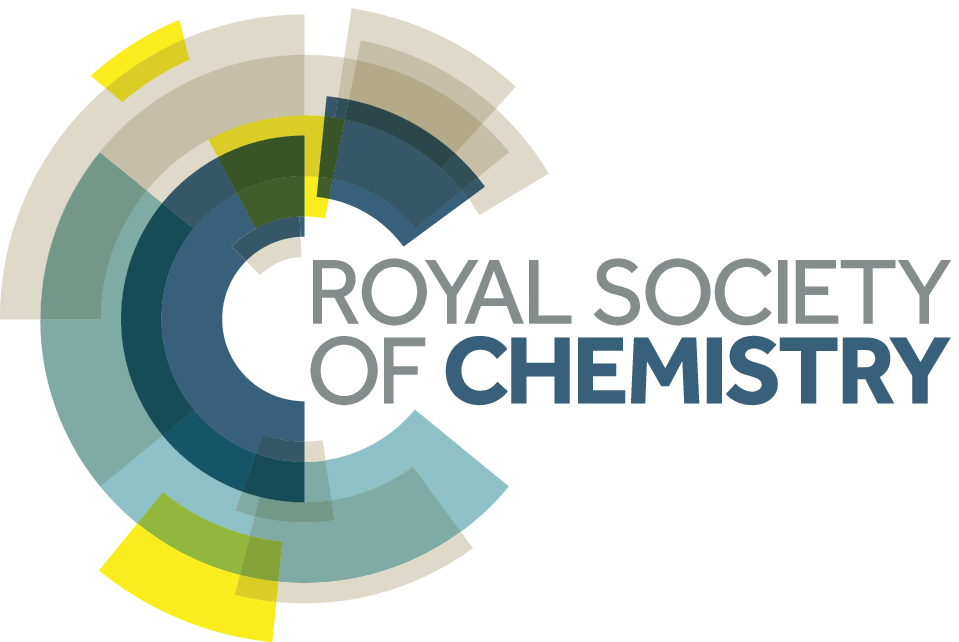}}
\renewcommand{\headrulewidth}{0pt}
}

\makeFNbottom
\makeatletter
\renewcommand\LARGE{\@setfontsize\LARGE{15pt}{17}}
\renewcommand\Large{\@setfontsize\Large{12pt}{14}}
\renewcommand\large{\@setfontsize\large{10pt}{12}}
\renewcommand\footnotesize{\@setfontsize\footnotesize{7pt}{10}}
\makeatother

\renewcommand{\thefootnote}{\fnsymbol{footnote}}
\renewcommand\footnoterule{\vspace*{1pt}%
\color{cream}\hrule width 3.5in height 0.4pt \color{black}\vspace*{5pt}}
\setcounter{secnumdepth}{5}

\makeatletter
\renewcommand\@biblabel[1]{#1}
\renewcommand\@makefntext[1]%
{\noindent\makebox[0pt][r]{\@thefnmark\,}#1}
\makeatother
\renewcommand{\figurename}{\small{Fig.}~}
\sectionfont{\sffamily\Large}
\subsectionfont{\normalsize}
\subsubsectionfont{\bf}
\setstretch{1.125} 
\setlength{\skip\footins}{0.8cm}
\setlength{\footnotesep}{0.25cm}
\setlength{\jot}{10pt}
\titlespacing*{\section}{0pt}{4pt}{4pt}
\titlespacing*{\subsection}{0pt}{15pt}{1pt}

\fancyfoot{}
\fancyfoot[LO,RE]{\vspace{-7.1pt}\includegraphics[height=9pt]{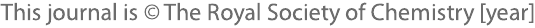}}
\fancyfoot[CO]{\vspace{-7.1pt}\hspace{13.2cm}\includegraphics{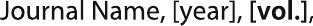}}
\fancyfoot[CE]{\vspace{-7.2pt}\hspace{-14.2cm}\includegraphics{head_foot/RF}}
\fancyfoot[RO]{\footnotesize{\sffamily{1--\pageref{LastPage} ~\textbar  \hspace{2pt}\thepage}}}
\fancyfoot[LE]{\footnotesize{\sffamily{\thepage~\textbar\hspace{3.45cm} 1--\pageref{LastPage}}}}
\fancyhead{}
\renewcommand{\headrulewidth}{0pt}
\renewcommand{\footrulewidth}{0pt}
\setlength{\arrayrulewidth}{1pt}
\setlength{\columnsep}{6.5mm}
\setlength\bibsep{1pt}

\makeatletter
\newlength{\figrulesep}
\setlength{\figrulesep}{0.5\textfloatsep}

\newcommand{\topfigrule}{\vspace*{-1pt}%
\noindent{\color{cream}\rule[-\figrulesep]{\columnwidth}{1.5pt}} }

\newcommand{\botfigrule}{\vspace*{-2pt}%
\noindent{\color{cream}\rule[\figrulesep]{\columnwidth}{1.5pt}} }

\newcommand{\dblfigrule}{\vspace*{-1pt}%
\noindent{\color{cream}\rule[-\figrulesep]{\textwidth}{1.5pt}} }

\makeatother


\newcommand{\argc}[1]{\left[#1\right]}
\newcommand{\arga}[1]{\left\lbrace #1\right\rbrace }
\newcommand{\argp}[1]{\left(#1\right)}
\newcommand{\valabs}[1]{\vert #1\vert}
\newcommand{\moy}[1]{\left\langle  #1 \right\rangle }
\newcommand{\moydes}[1]{\overline{#1}}
\renewcommand{\vec}[1]{{\bf #1}}
\newcommand{\meevid}[1]{\textcolor{VioletRed}{#1}}
\newcommand{\KM}[1]{\textcolor{blue}{#1}}
\newcommand{\comment}[1]{\textcolor{OliveGreen}{#1}}



\twocolumn[
  \begin{@twocolumnfalse}
\vspace{3cm}
\sffamily
\begin{tabular}{m{4.5cm} p{13.5cm} }

\includegraphics{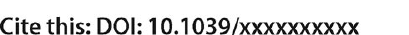} & \noindent\LARGE{\textbf{Residual stress in athermal soft disordered solids: insights from microscopic and mesoscale models}} \\
\vspace{0.3cm} & \vspace{0.3cm} \\

 & \noindent\large{Vishwas V.Vasisht$^{\ast}$\textit{$^{a,b}$}, Pinaki Chaudhuri\textit{$^{c}$} and  Kirsten Martens\textit{$^{a}$}} \\
\includegraphics{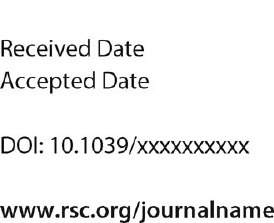} & \noindent\normalsize{

In soft amorphous materials, shear cessation after large shear deformation leads to structures having residual shear stress. The origin of these states and the distribution of the local shear stresses within the material is not well understood, despite its importance for the change in material properties and consequent applications. In this work, we use molecular dynamics simulations of a model dense non-Brownian soft amorphous material to probe the non-trivial relaxation process towards a residual stress state. We find that, similar to thermal glasses, an increase in shear rate prior to the shear cessation leads to lower residual stress states. We rationalise our findings using a mesoscopic elasto-plastic description that explicitly includes a long range elastic response to local shear transformations. We find that after flow cessation the initial stress relaxation indeed depends on the pre-sheared stress state, but the final residual stress is majorly determined by newly activated plastic events occurring during the relaxation process. Our simplified coarse grained description not only allows to capture the phenomenology of residual stress states but also to rationalise the altered material properties that are probed using small and large deformation protocols applied to the relaxed material.

} \\

\end{tabular}

 \end{@twocolumnfalse} \vspace{0.6cm}

  ]
 


\renewcommand*\rmdefault{bch}\normalfont\upshape
\rmfamily
\section*{}
\vspace{-1cm}


\footnotetext{\textit{$^{*}$~E-mail:} vishwas@iitpkd.ac.in}
\footnotetext{\textit{$^{a}$~Univ. Grenoble Alpes, CNRS, LIPhy, 38000 Grenoble, France.}}
\footnotetext{\textit{$^{b}$~ Dept. of Physics, Indian Institute of Technology, Palakkad.}}
\footnotetext{\textit{$^{b}$~The Institute of Mathematical Sciences, Taramani, Chennai 600113, India.}}

\section{Introduction}
\label{section-intro}
Soft glassy materials are ubiquitous in our everyday life and find their application in various domains, such as food science, pharmaceutical engineering or medical applications \cite{Bonn_RMP_2017, joshi2018yield}. Being easier to handle and visualise in experimental setups, these materials have become a playground for analysing diverse mechanical response of amorphous solids in general, since many features are common to both hard and soft materials.

Despite a large body of empirical knowledge, the fundamental understanding of the mechanical behaviour of dense disordered materials is still a work in progress \cite{Bonn_RMP_2017, joshi2018yield, coussot2014yield, nicolas2018deformation, rodney2011modeling}. Besides many interesting properties in the deformation process, like complex yielding phenomena and non-trivial rheology with possible flow localization, these materials also display fascinating behaviour once they are let to relax. The most striking observation is that they can hold internal stresses even long after the cessation of their mechanical stimulus \cite{Withers_RoPP_2007}. Such locked-in stresses, known as residual stress, can occur from diverse procedures, e.g. at the end of thermal quenches (e.g., Rupert's drop \cite{brodsley1986prince, kooij2021explosive}), due to chemical processes (e.g., Tempered glass \cite{garfinkel1970ion}) or by the cessation of some external mechanical drive \cite{withers2001residual}. Understanding how such residual stresses are created is of fundamental interest, within the broader framework of identifying specific microscopic processes responsible for the observed mechanical properties of amorphous systems. Further, knowing how such stresses build up during the formation of the material, is also of practical importance, since there can be both detrimental or beneficial aspects depending upon the specific functional aspects of the material under consideration \cite{Withers_RoPP_2007}. For example, in many cases, residual stresses can lead to materials being more prone to fracture \cite{reiter2005residual}. On the other hand, such stresses also allow some materials to be resistant or control the extent of failure \cite{green1999crack, zhang2006making}. Therefore, the ability to control the extent of residual stresses is significant for the design of new material properties, and thus there is a need for better understanding of the processes involved.

In the context of soft amorphous materials, various studies have  recently extensively explored  the occurrence of residual stresses, via switching off the flow driven by an external shear-rate \cite{Bonn_RMP_2017, chung2006microscopic, bandyopadhyay2010stress, negi2009dynamics, Negi_JoR_2010, Ballauff_PRL_2013, mohan_prl_2013, mohan_JoR_2015, Moghimi_SM_2017}. For the case of non-Brownian systems (e.g. emulsions), upon switch off, the shear stress, accumulated hitherto during flow, decays rapidly followed by a slower relaxation to an eventual plateau, i.e. a stuck state with a well-defined residual stress is obtained \cite{mohan_prl_2013, mohan_JoR_2015}. Further, a scaling relation between initial stress, at the point of shear switch off, and the corresponding locked-in stresses was revealed. For the case of Brownian systems (e.g. colloidal glasses, gels), the eventual plateau is not observed. Rather, a power-law decay in stress continues to occur, due to aging processes characteristic to glasses \cite{Ballauff_PRL_2013}. Mode coupling calculations for model glassy systems also report similar behaviour \cite{Fritschi_SM_2014}. 

In this work, we probe the origin of residual stresses in model athermal amorphous systems, via a combination of microscopic and mesoscale simulations. First, using molecular dynamics simulations of a model dense non-Brownian soft amorphous material, we demonstrate that dynamically arrested states having finite residual stresses are obtained upon the cessation of an applied shear, consistent with experimental observations. As is also demonstrated via experiments, the final residual stress depends upon the strain-rate at which the system was being driven, prior to the switch-off of shear;  larger shear-rate for the initially flowing states lead to eventual arrested states having lower residual stress. We rationalise our findings using a mesoscopic elasto-plastic description that explicitly includes a long range elastic response to local shear transformations. 

In a serious of works, it has been shown that these type of elasto-plastic descriptions account for various phenomena in the deformation process of dense disordered media. Since their introduction for the description of local plasticity \cite{argon1979plastic, baret2002extremal, picard2004elastic}, they have been used in a variety of contexts. Notably they have extensively used to study critical fluctuations in the yielding transition \cite{lin2014scaling, liu2016driving, Budrikis2017plastic, karimi2017inertia, ferrero2019criticality} as well as in the vicinity of a finite shear rate critical point \cite{le2019criticality, le2020giant}. Further, they have been useful for the understanding of strain localization \cite{talamali2012strain, jagla2007strain} and permanent shear banding \cite{martens2012spontaneous, barbot2020rejuvenation}, and also creep phenomena \cite{merabia2016thermally, liu2018creep}. And more recent developments have been successful to establish a more quantitative link between these coarse grained elasto-plastic descriptions and the microscopic dynamics measured in particle-based simulations \cite{puosi2014time, puosi2015probing, albaret2016mapping, patinet2016connecting, liu2021elastoplastic}. 

Our elastoplastic model, used in this study, exhibits all the phenomenological observations related to residual stress measurements, observed in experiments and our microscopic simulations. Using the ability to track local plastic events in such elastoplastic models, we provide a semi-analytical analysis of the formation of residual stresses after the flow cessation. Our significant finding is that the final residual stress is majorly determined by newly activated plastic events occurring during the relaxation process, after the shear switch-off. Finally, using both the microscopic and mesoscale model, we illustrate that states having less residual stresses are more rigid, which becomes evident via the observed transient response to applied external shear. 


The paper is organised as follows. After the introductory discussion in Section 1, we elaborate the microscopic and mesoscale models that we have studied in Section 2, along with the the methods involved in the numerical simulations. In Sections 3 and 4, we report and discuss the measurement of residual stresses in the microscopic and mesoscale models. In Section 5, we provide the semi-analytical analysis rationalising the observations regarding the residual stress measurements. In Section 6, we illustrate some of the mechanical characteristics of the residual stress states. Finally, in Section 7, we provide a concluding discussion.

\section{Model and protocol}
\label{section-model}
\subsection{Molecular dynamics simulations}

In our study, the dense amorphous solid is modelled as a non-Brownian suspension of soft repulsive spheres, at a volume fraction $\phi \approx 70\%$, consisting of $97556$ particles, with repulsive effective interactions mimicked via a truncated and shifted Lennard-Jones potential \cite{WCA_JCP_1971} given by $U(r) = 4\epsilon \left [(a_{ij}/r_{ij})^{12} - (a_{ij}/r_{ij})^{6} \right ] + \epsilon$, for $r_{ij} \le 2^{1/6} a_{ij}$, else $U(r_{ij})=0$. Here $\epsilon$ is the unit energy in the simulations, $r_{ij}$ being the center to center distance between the particle $i$ and $j$ and $a_{ij}=0.5(a_i+a_j)$, with $a_i$ and $a_j$ being the diameter of particles $i$ and $j$ respectively. The diameters of the particles are drawn from a Gaussian distribution with variance of $10\%$, whose mean is used as unit length $a$. We prepare the initial samples by quenching  high temperature liquid states ($T=5\epsilon/k_B$) to low temperature ($T=0.01\epsilon/k_B$) using a NVT Molecular Dynamics protocol at a fixed cooling rate $\Gamma$ ($10^{-4} \epsilon/k_B\tau_0$). Each sample is subsequently brought to the closest energy minimum and to $k_B T/\epsilon \approx 0$ via energy minimization. Under athermal conditions, these samples are subjected to a shear rate $\dot{\gamma}$ using Lees-Edwards boundary conditions (LEBC) and solving the following dissipative particle dynamics (DPD) based equations of motion.
\begin{equation}
	m \frac{d^2\vec{r}_i}{dt^2} = - \zeta_{DPD} \sum_{j(\ne i)} \omega(r_{ij}) (\hat{r}_{ij}.v_{ij})\hat{r}_{ij}  - \triangledown_{\vec{r}_i} U
\end{equation}
\noindent where $m$ is the mass of the particle, the first term in the right hand side (RHS) is the damping force which depends on the damping coefficient $\zeta_{DPD}$. We have taken $\zeta_{DPD}=1.0$ that guarantees minimum inertial effects \cite{vasisht_PRE2_2020}. The relative velocity $\vec{v}_{ij} = \vec{v}_j - \vec{v}_i$ is computed over a cut-off distance $r_{ij} \le 2.5 a_{ij}$, with the weight factor $\omega(r_{ij})=1$. The second term in the RHS is the force due to interactions between particles. In all our simulations [x,y,z] dimensions refers to flow, gradient and vorticity directions respectively. The shear stress is computed from the complete virial stress tensor as $\sigma_{xy} \equiv \sigma = \frac{1}{V} \sum_{i} \sum_{j>i} x_{ij} f^y_{ij}$, where V($=lx*ly*lz$) is the volume of the system, $x_{ij}$ represents x-component of the distance between particle $i$ and $j$ and $f^y_{ij}$ is the y-component of  force on the particle $i$ due to $j$. 

Also, for the residual stress states, we compute the storage modulus by performing small strain oscillatory shear simulations via the application of a shear strain $\gamma(t) = \gamma_0 sin(\omega t)$, with a strain amplitude of $\gamma_0 = 1\%$. Computing the stress response for frequencies $\omega$, from the steady state regime we extract visco-elastic coefficient using 
\begin{equation}
     G'(\omega, \gamma_0) = \frac{\omega}{\gamma_0 \pi} \int_{t_0}^{t_0 + 2\pi/\omega} \sigma_{xy}(t) sin(\omega t) dt
\end{equation}
\noindent The storage modulus is obtained using  $G'= G'(\omega \rightarrow 0, \gamma_0)$.

All the microscopic simulations are done using LAMMPS \cite{plimpton1995fast}, with a modification to handle polydispersity in size.

\subsection{Mesoscopic simulations}
The mesoscopic simulations we perform in this study are based on an elasto-plastic model for a yield stress material under steady shear at an applied shear rate $\dot{\gamma}$. The model consists of a regular square lattice with each lattice site representing an elasto-plastic element holding a shear-stress variable evolve according to

\begin{equation}
\frac{\partial \sigma_i}{\partial t} = \mu \dot{\gamma} + \mu \sum_j G_{ij} \frac{\partial {\gamma}^{pl}_j}{\partial t}
\end{equation}

\noindent where $\sigma_i$ is the scalar shear stress component at a site i, $\mu$ is the material dependent shear modulus, and hence $\mu \dot{\gamma}$ is the elastic contribution to the evolution of stress. The second term on the right hand side (RHS) accounts for the change in stress due to local plastic yielding. Here $\partial {\gamma}^{pl}_j/\partial t$ is the rate of plastic strain deformation and $G_{ij} \propto \cos{(4\theta)}/r^2$ is the Eshelby kernel which accounts for the stress redistribution due to a plastic event. The sum over $j$ in the second term on the RHS represent the rate of local elastic deformation associated with the response to a plastic deformation on a distant site j. At the plastically deforming site, the relaxation dynamics is modeled as Maxwellian visco-elastic relaxation $\partial {\gamma}^{pl}_j/\partial t = (1/\mu \tau) n_j \sigma_j$, where $\tau$ is the characteristic time for stress release in a plastic phase, a system property which depends on the volume fraction $\phi$, $n_j$ is the local state variable which refer to the local activity. $n_j=1$, if the site is in the plastic phase, otherwise $n_j=0$. Similar to Ref. \cite{martens_SM_2012}, a stochastic dynamics is followed for the evolution of state variable $n$. If at a given site $i$ the system has $\sigma_i < \sigma_y$, with site variable $n_i=0$, according to equation 1, the stress changes due to elastic contribution ($\mu \gamma$). If the local stress $\sigma_i$ is greater than the local yield stress $\sigma^i_y$, the $n_i$ transform from $0$ to $1$ within a time period of $\tau_{pl}$. Once the system yields, the stress relaxation has a local contribution as well in the  second term in the equation 1. Within the plastically active element the stress relaxes as $e^{-g t/ \tau}$ during a typical time $\tau_{res}$, where g is numerical value of the propagator at site i, $G_{ii}$, that insures mechanical equilibrium. In two dimensions in the large system limit the value of g tends to $g\approx0.57$. 

In our model the active state ($n_j=1$) relaxes back to $0$ with a rate $\tau_{res}^{-1}$, where $\tau_{res}$ is the typical local restructuring time of the material to regain its local elastic properties after a local yield event. In our protocol the time scales $\tau_{pl}$ and $\tau_{res}$ are typically smaller than the driving time scale $\tau_{a} = \dot{\gamma}^{-1}$. The distribution of local yield stress threshold values follows from an argument wherein yielding is considered as a jump in the potential energy landscape (PEL) with energy barrier $E_y = \sigma^2_y/4\mu$. The distribution is given by $P(E_y)= \Theta(E_y-E^{min}_y) \lambda exp(-\lambda (E_y - E^{min}_y$)), where $E^{min}_y$ term considers only larger jumps ignoring the small jumps within the metabasins of PEL, the parameters $\lambda$ and $E_{min}$ determine the average yield strain $\gamma_y$. We choose $E_{min}=0.001$ and $\lambda=2015$ such that we obtain strain overshoot in the shear start-up regime while performing re-shear simulations, reaching a residual stress state.

We emphasise here that our objective is not develop a mesoscale model of the microscopic model that we study. Rather, the objective is to demonstrate that the mesoscale model qualitatively reproduces the phenomenology observed in the microscale model and thereafter use the simplicity of the mesoscale model to gain insights into the physical processes involved in the creation of residual stresses in amorphous solids.

\begin{figure}[h!]
    \includegraphics[width=0.45\textwidth]{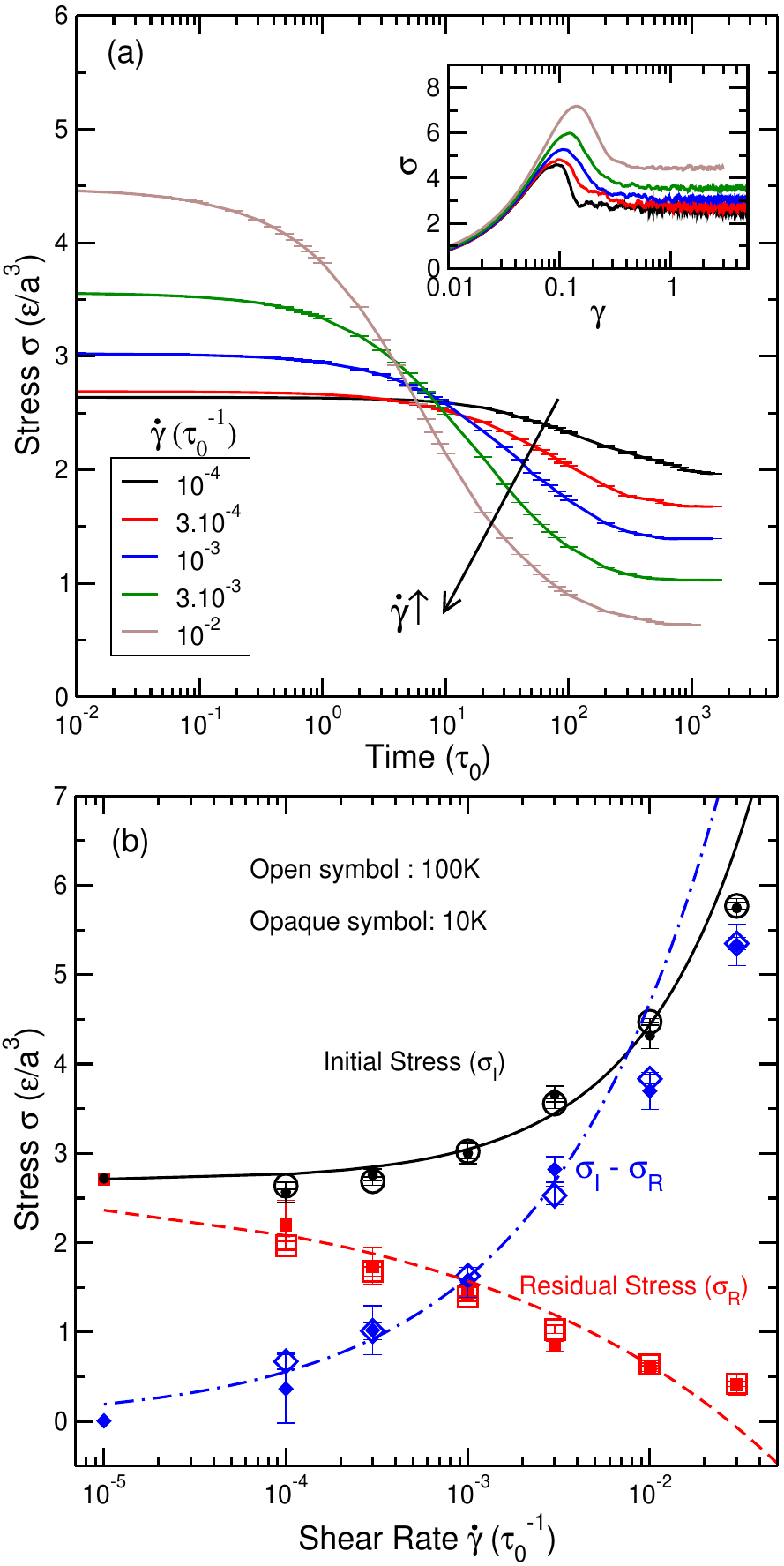}
    \caption{{\bf Microscale particulate model.} (a) Relaxation of shear stress ($\sigma$) when the external drive (imposed shear-rate) is switched off, while in steady state flow. The arrow shows the direction of increasing shear-rate ($\dot{\gamma}$). (inset) Load curve ($\sigma$ vs. strain $\gamma$) showing the start-up to steady state regime, for different shear rates, prior to switch-off. (b) The stress $\sigma_I$ at the shear switch-off (red square), the residual stress $\sigma_R$ reached at the end of stress relaxation upon switch-off (black circle) and $\Delta \sigma = \sigma_I-\sigma_R$ (blue diamond), shown for two different system sizes (see labels) as a function of the imposed shear rate.}
    \label{Fig1}
\end{figure}

\section{Approach to residual stress states in microscopic simulations} 

In this section, we discuss our microscopic simulation results related residual stress states and it's dependence on the shear rate. 

As discussed above, we initially impose an external shear deformation at a chosen shear rate $\dot{\gamma}$ till the system reaches a steady state regime. The corresponding variation of shear stress ($\sigma$) with shear strain ($\gamma$), for different imposed $\dot{\gamma}$ is shown in the inset of Fig.\ref{Fig1}(a). Once in steady state, we switch off the external shear and allow the accumulated stress to relax while keeping the strain fixed. The stress decreases with time and is observed to reach a finite saturation value, which is referred to as the residual shear stress $\sigma_R$ of the system. To obtain good statistics for $\sigma_R$, we not only perform three different independent runs, but also relax the system from 5 different flow cessation points for each run. Further, we check for finite size effects by probing the behaviour for two different system size, viz. N=91556 and N=10976. 

The relaxation of stress, for the different imposed shear-rates, is shown in the main panel of Fig.\ref{Fig1} (a). At a given shear rate, upon flow cessation, the stress relaxes steadily at small times, followed by a dramatic decrease and eventual arrest to a fixed value. Also, the timescale for the stress relaxation
is fastest for the largest $\dot{\gamma}$ and slows down with increasing shear-rate. These features are similar to previous reported work on emulsions \cite{mohan_prl_2013} and attractive gels \cite{Moghimi_SM_2017}, and unlike thermal glasses where aging continues after the initial stress relaxation \cite{Ballauff_PRL_2013}. The dependence of $\sigma_R$ upon the initially imposed $\dot{\gamma}$ is shown in the Fig. \ref{Fig1} (b). Consistent with previous reports, $\sigma_R$ decreases with increasing $\dot{\gamma}$. Also shown in  Fig. \ref{Fig1} (b), are the values of the shear stress at flow cessation  $\sigma_I$ and the difference between the stress in flow and and the residual stress at arrest, $\Delta \sigma = \sigma_I - \sigma_R$. The shear rate dependence of $\sigma_I$ is well described by the Herschel-Bulkley function $\sigma_y + \kappa \dot{\gamma}^n$, as expected, where the dynamic yield stress $\sigma_y = 2.7 \epsilon/a^3$ and the exponent $n=0.7$. In the limit of vanishing $\dot{\gamma}$, as the stress in the system reaches the dynamic yield stress, $\Delta \sigma$ approaches to zero, i.e. the residual stress value coincides with the dynamical yield stress. On the other hand, with the increase in shear rate where the $\sigma_R$ decrease, $\Delta \sigma$ increases following a power-law profile (solid lines in the Fig. \ref{Fig1} (b)). Finally, note that, we find that there is very little variation observed with changing system size (as seen from the open symbols with the opaque symbols in Fig. \ref{Fig1} (b)). 

In the case of thermal glasses, Ballauff {\it et al.}   \cite{Ballauff_PRL_2013} associate the residual stresses with the long-lived memory effects of the pre-sheared glasses which introduce supra-caging lengthscale in the system. Long time stress relaxation are associated with ageing effects. In the case of non-Brownian colloidal gel like systems, Mohan {\it et al.}\cite{mohan_prl_2013} find  microstructural signature of the stress relaxation which associate with different structural rearrangements. An initial rapid relaxation is associated with unjamming process where in a ballistic motion is observed and long time slow relaxation is associated with cage breaking process. 
In this work we address the questions related to approach towards residual stress state using a elasto-plastic model based approach using mesoscopic simulations. 

\begin{figure}[h!]
\includegraphics[width=0.45\textwidth]{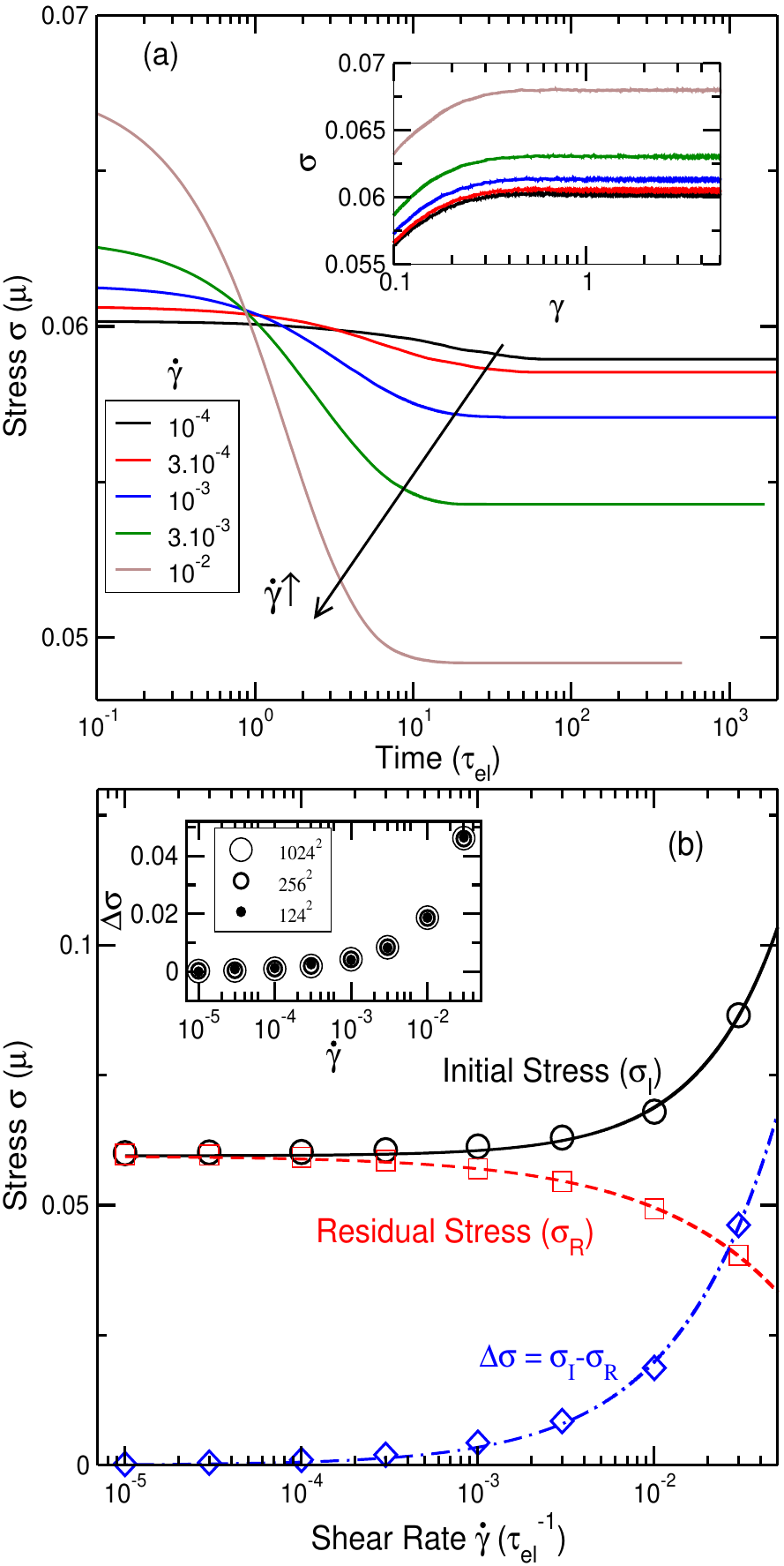}
\caption{{\bf Mesoscale model.} (Top) Relaxation of shear stress ($\sigma$) when the external drive (imposed shear-rate) is switched off, while in steady state flow. The arrow shows the direction of increasing shear-rate ($\dot{\gamma}$). Overall behaviour is similar to the response observed in microscale simulations as shown in Fig.\ref{Fig1}. (inset) Load curve ($\sigma$ vs. strain $\gamma$) showing the start-up to steady state regime, for different shear rates, prior to switch-off. (Bottom) The stress $\sigma_I$ at the shear switch-off (red square), the residual stress $\sigma_R$ reached at the end of stress relaxation upon switch-off (black circle) and $\Delta \sigma = \sigma_I-\sigma_R$ (blue diamond), shown for $N=1024^2$ as a function of the imposed shear rate (system size dependence is shown in the inset). Again, overall behaviour similar to what is observed in particulate model; see Figs.\ref{Fig2}. }
\label{Fig2}
\end{figure}

\section{Approach to residual stress states in mesoscale simulations} 

Similar to particulate simulations, in the mesoscale simulations, upon flow cessation, the sheared state relaxes to a state having finite residual stress (Fig. \ref{Fig2}(a)). The start-up to steady state evolution of the stress is shown in the inset of top panel of Fig.\ref{Fig2} (a), and the relaxation of the shear stress, after the flow cessation, is shown in main panel Fig.\ref{Fig2}(a). In the mesoscale model, the flow cessation is mimicked by turning off the first term ($\mu \dot{\gamma}$) in the RHS of Equation 1. The variation of the $\sigma_R$ with the shear-rate during flow, $\dot{\gamma}$, is shown in Fig.\ref{Fig2} (b), along with the stress ($\sigma_I$) at which the switch off occurs, and the corresponding stress difference $\Delta{\sigma} = \sigma_I - \sigma_R$. The qualitative behaviour of all these quantities is very similar to what has been observed for the particulate system, as shown in Fig.\ref{Fig1}. The data for $\sigma_I$ can also be fitted with a Herschel Bulkley law, obtaining $\sigma_y=0.06$ and $n=0.8$. Again, the system size dependence for $\Delta{\sigma}$ does not show much variation (Fig.\ref{Fig2} (b) inset). Thus, overall, the mesoscale model can successfully reproduce all the qualitative features related to the observations on residual stress that we obtain in the microscopic model. 

By construction, the mesoscale simulations provides easy access to local yielding events or plastic events and its spatio-temporal evolution and hence we utilize  these simulations to analyse the process of approaching the locked-in or residual stress state and it's dependence on flow rate prior to shear switchoff.

\begin{figure*}
\begin{center}
\includegraphics[width=1\textwidth]{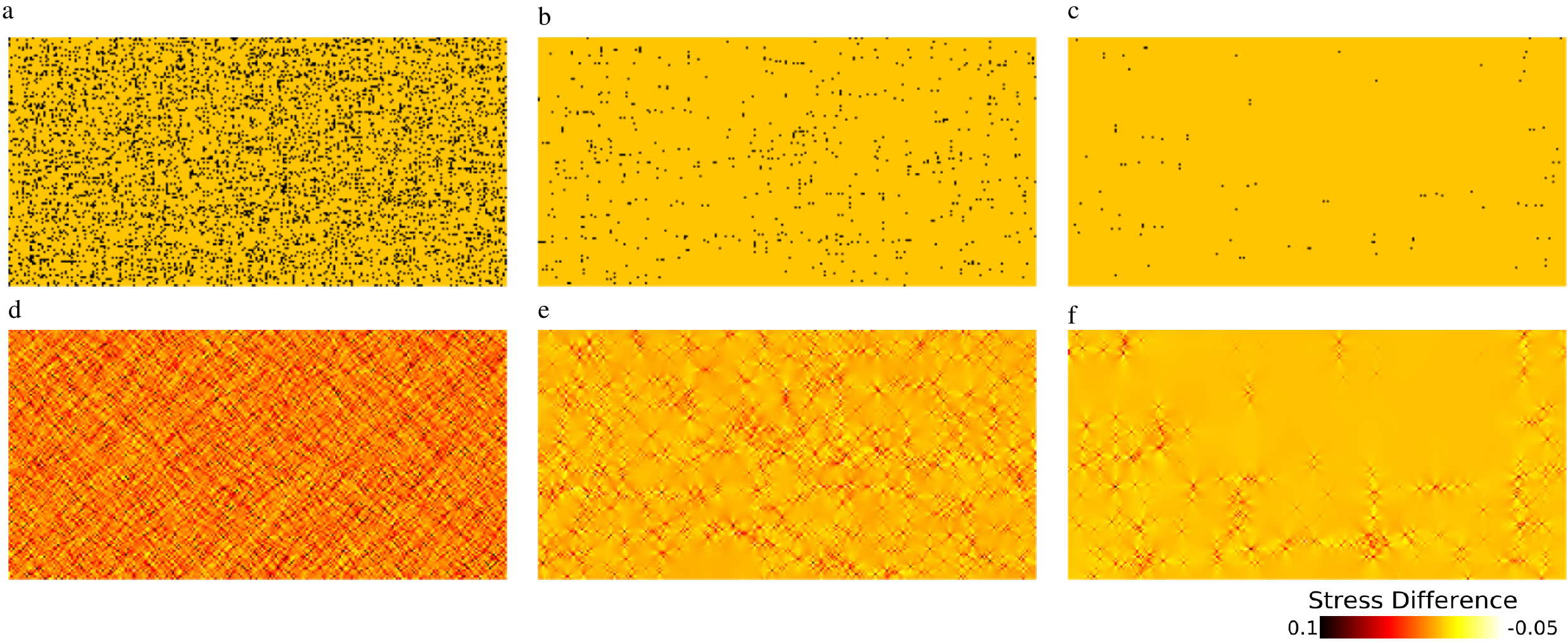}
\caption{{\bf Mesoscale model.} (Top) Maps of state of activity at the point of shear switch-off, for different driving rates (a) $10^{-2}$, (b) $10^{-3}$, (c) $10^{-2}$. Black color correspond to active or plastic site ($n_i$=1) and Yellow correponds to $n_i$=0. (Bottom) Corresponding maps of local stress difference between initial 
and final stress, $\delta\sigma_i$ for the corresponding shear rates in top. In all cases, only half of the simulation domain is shown for clarity purposes.}
\label{Fig3}
\end{center}
\end{figure*}

\vspace{-4mm}
\subsection{Mesoscale stress relaxation maps}

To begin with we study, in the mesoscale simulations, how well the the configurations at the flow cessation regulate the final residual state. 

In Fig. \ref{Fig3} (a-c) we show the local activity map of the initial configurations for three different shear rates. In the maps the yielded sites ($n_i=1$) (hereby termed active) is represented by black color and the yellow depicts the elastic sites ($n_i=0$) (henceforth termed inactive). As expected, with the decrease in shear rates (see Fig. \ref{Fig3} (a-c)), the number and also the spatial density of active sites decreases. Naively, one expects that when the shear is switched off, these active mesoblocks release their stress and become inactive, leading to the residual stress states. 
However spatial maps of local stress difference between initial flowing and final arrested states, viz. $\Delta{\sigma}_i = (\sigma_I - \sigma_R)_i$ (see Fig. \ref{Fig3} (d-f)) tell a different story.  We observe that $\Delta{\sigma}_i$ is finite in sites beyond the initial active sites, which is very evident even at the low shear limit, see Fig.\ref{Fig3}(f). This clearly indicates that  the stress relaxation process, upon flow cessation, involves more sites than the ones which were active
at the time of shear switch-off, and spatial scale of this become more and more extensive with increasing pre-sheared $\dot{\gamma}$.

\subsection{Activity statistics in mesoscale model}

Following the above analysis vis-a-vis the stress difference maps, we now proceed to quantify our observations. 

\begin{figure}[th!]
\includegraphics[width=0.45\textwidth]{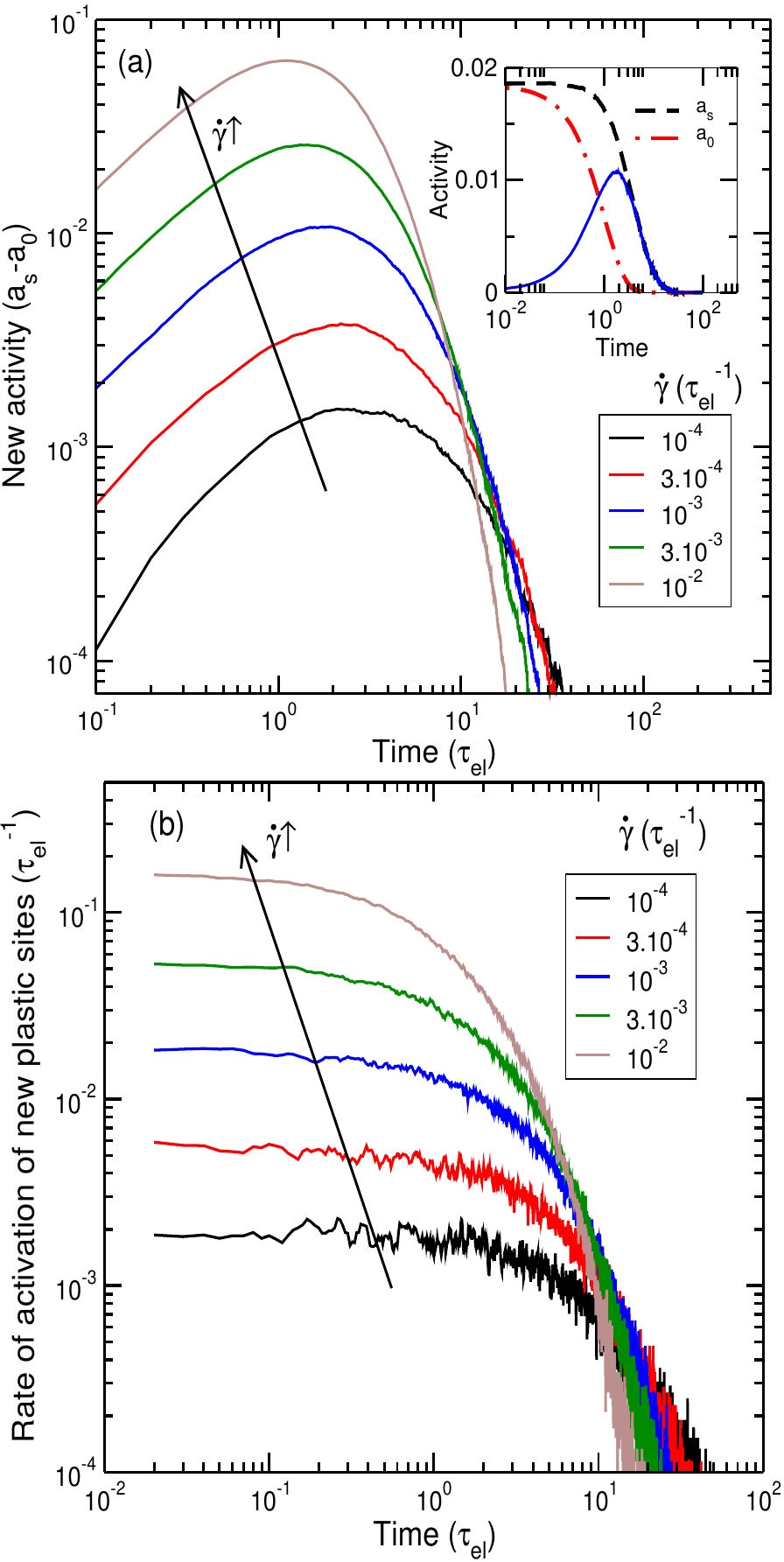}
\caption{{\bf Mesoscale model.} (Top) For different imposed shear-rates during the steady flow, the variation in the number of newly activated sites (see text for discussion) as a function of time, during stress relaxation after shear switch-off. Arrow shows the direction of increasing shear-rate. (inset) Variation of total number of activated sites (in black) with time, during the relaxation, as well as the decay of pre-existing activated sites within the configuration at switch-off (in red), apart from the newly activated sites (in blue) (for $\dot{\gamma}=10^{-3}$). (Bottom) Rate of activation of new plastic sites as a function of the time during the stress relaxation following switch-off. Arrow shows the direction of increasing shear-rate.}
\label{Fig4}
\end{figure}

We start by identifying the active sites at the time of shear switch-off and compute the fraction of such sites in the system, $a_0$. During the stress relaxation, we monitor these identified sites and check how many remain active at any time instant, and thereby arrive at the time evolution of $a_0(t)$. At any time instant, we also count the overall fraction of observed active sites, $a_s(t)$. The difference, $a_s(t)-a_0(t)$,
gives the fraction of sites that have been newly activated.  In the inset of Fig. \ref{Fig4} (a), for system size of $1024^2$, we show how $a_0$, $a_s$ and the difference $a_s-a_0$ evolve in time, for the case of relaxation from a state that was being sheared at $\dot{\gamma}=10^{-3}\tau^{-1}$. 
We find that from very early times,  $a_s(t) > a_0(t)$, indicating the formation of new active sites. This is evident from the behaviour of $a_s(t)-a_0(t)$, which grows with time as relaxation proceeds, goes through a maximum and then eventually decays. Further, we also note that $a_0(t)$, i.e. the original active sites, deplete very soon, whereas $a_s(t)$ takes longer time to die out, implying that the late time relaxation is due to the decay of the newly activated sites. 

In the main panel of Fig. \ref{Fig4} (a), we show the time evolution of $a_s(t)-a_0(t)$ for different values of shear-rate at which the system was flowing prior to shear switch-off. In all cases, we observe the non-monotonic growth and decay of the newly activated sites. The peak in $a_s(t)-a_0(t)$ is highest for the case of switch-off from the largest shear-rate and decreases with decreasing shear-rate. The presence of larger fraction of active sites at the point of switch off from a flow having large shear-rate, is likely to lead to generate more active sites and therefore larger peak.  Next, as is the case with residual stress, the eventual decay in the number of active sites is fastest for the case of shear switch-off from the flowing state having the highest shear-rate. 

In Fig. \ref{Fig4} (b), we show the rate of production of the new active sites, directly measured from the mesoscale simulations. As is evident, the rate is higher for the case of the flowing state having largest shear-rate and the decay is also fastest, as pointed out above. 

Both Fig. \ref{Fig4} (a) and (b) show that beyond a certain time 
the system no longer creates new active sites (or no new local yielding events are observed) and the existing active sites continue to relax stress. 

Even though identifying plastic sites in molecular simulation is a challenging task, using the knowledge of mean square displacement and the measure of non-affine displacements $d^2_{min}$ \cite{}, we confirm that a similar observations are made in the particulate system. 

Hence the emerging scenario is that the residual stress obtained at the end of stress relaxation in these amorphous states, depends not only on the state of activity in the initial sheared state but also on the dynamics during the stress relaxation. We utilize these observations to provide a semi-analytical rational for the magnitude of residual stresses obtained from the mesoscale simulations.

\section{Semi-analytical analysis of residual stress}

We propose a simple analytical calculation which utilizes the observations we discussed in the previous section to predict the residual stress values, with inputs from mesoscale simulations. 

In the calculation, we define the residual stress $\sigma_R$ as the stress retained in the system after subtracting from the initial stress $\sigma_I$ the stress relaxed through the process of local yielding. In our mesoscopic model,  every time a site yields, the stress relaxes exponentially over a certain time window as $\exp{(-gt/\tau)}$. Given that the local yielding process is stochastic in nature, we will have a distribution the duration of events $P(\tau_{res})$. For an initial sample (at the flow cessation), it is plausible that fraction of the active sites might have yielded at earlier times and hence would have already relaxed to an extent. Assuming a uniform distribution of active sites, we can simply write residual stress as
\begin{equation}
    \sigma_{R} = \sigma_{I} - \int_0^1 dx P_1(x)  \int_0^{\infty} d\tau_{res} P(\tau_{res}) \int_0^{x\tau_{res}} dt \frac{n^{ini}_a}{N^2} \left\langle \sigma^{ini}_a \right\rangle \exp(-gt/\tau)
\end{equation}
\noindent where $(n^{ini}_a/N^2)$ gives the average fraction of active sites in the system when the shear is switched off  and $\left< \sigma^{ini}_a \right>$ is the average stress per active site at that time. The product of the two gives the total plastic stress which will eventually relax. We input this quantity from the mesoscale simulations. With the simplified expression for fraction of relaxation before shear cessation of the already active sites with $P_1(x)=1$ being the uniform distribution on the interval zero to one and $P(\tau_{res})=(1/\tau_{res})\exp{(-\tau_{res})}$ describing the stochastic elastic recovery, we obtain the residual stress as
\begin{equation}
    \sigma_{R} = \sigma_{I} - \frac{n^{ini}_a}{N^2}\left\langle \sigma^{ini}_a \right\rangle \tau \tau_{res}\left( \frac{\tau_{res}\log(\tau_{res})}{g^2\tau^2} - \frac{\tau_{res}\log(\tau_{res}+g\tau) - g\tau}{g^2 \tau^2} \right)
\end{equation}
\noindent Taking $\tau=1$ and $\tau_{res}=1$ as in the simulations we would like to compare to, this expression simplifies to
\begin{equation}
    \sigma_R = \sigma_I - \frac{n^{ini}_a}{N^2}\left\langle \sigma^{ini}_a \right\rangle \left( \frac{g - log(1+g)}{g^2} \right )
\end{equation}

\begin{figure*}
\includegraphics[width=1\textwidth]{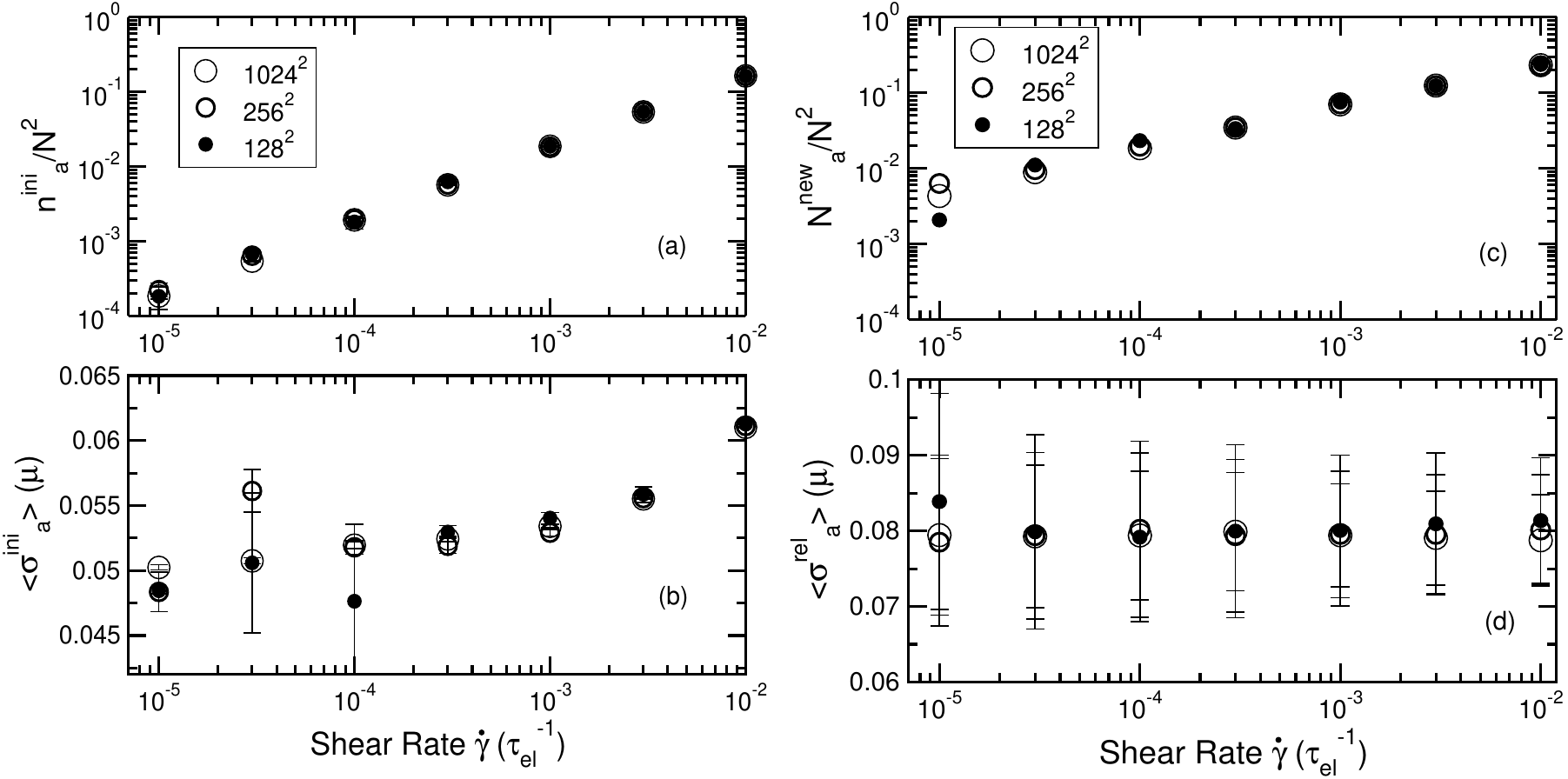}
\caption{{\bf Mesoscale model.} (Left) (top) Fraction of active sites, $n^{ini}_a/N^2$ in the flowing state, at the point of shear switch-off, and (bottom) corresponding  average stress per active site, $\langle \sigma^{ini}_a\rangle$, as a function of shear rate, shown for different system sizes. (Right) (top) The cumulative fraction of active sites, $N^{new}_a/N^2$, computed from the flow cessation to the eventual relaxed arrested state, and (bottom) associated stress per active site, $\langle \sigma^{rel}_a \rangle$, averaged over the whole range of relaxation, as a function of shear rates, shown for different system sizes.}
\vspace{5mm}
\label{Fig5}
\end{figure*}

\begin{figure}[b!]
\includegraphics[width=0.45\textwidth]{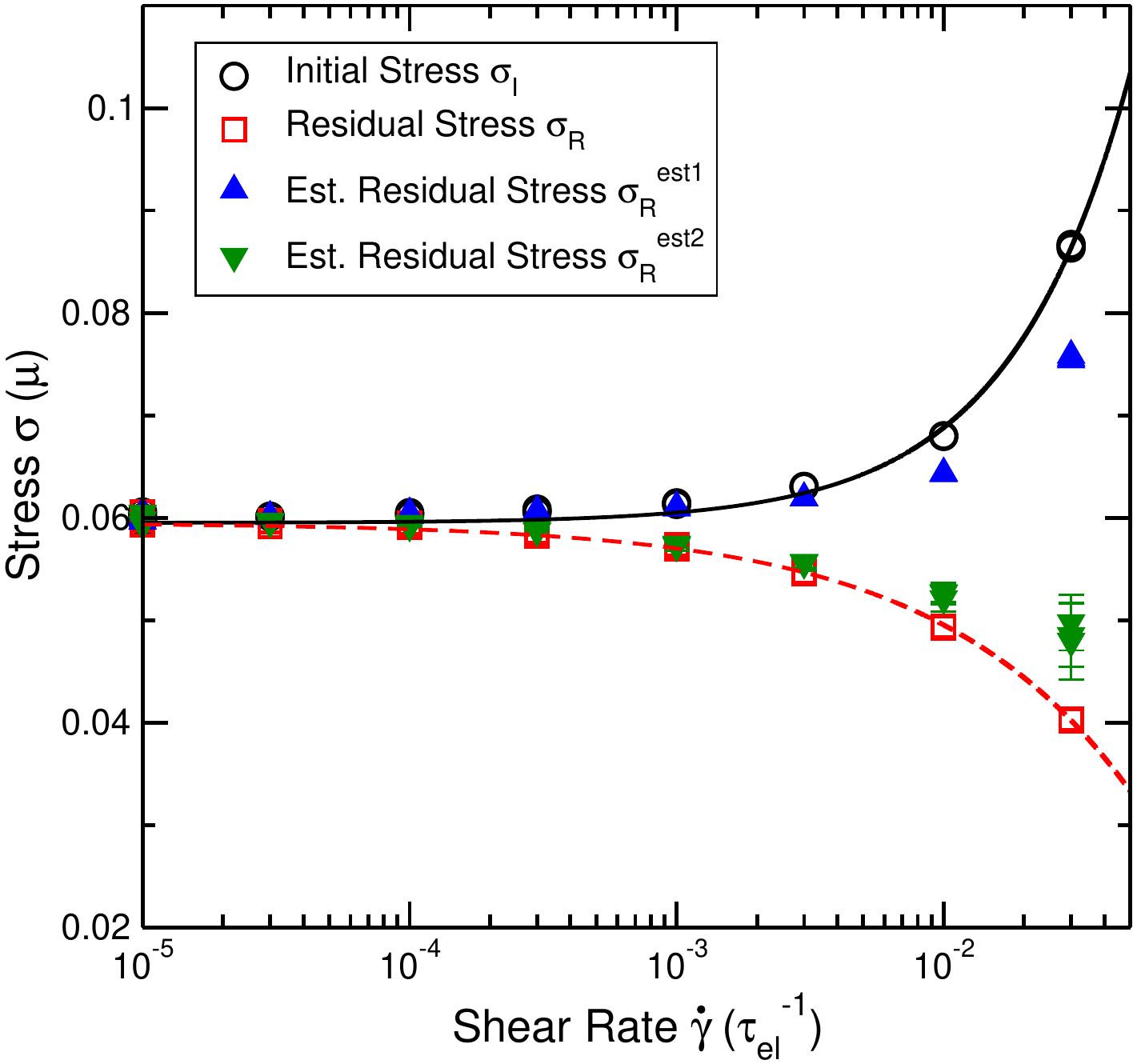}
\caption{{\bf Mesoscale model.} Comparison of residual stress values measured from simulations and estimated from semi-analytical calculations (see main text).}
\label{Fig6}
\end{figure}

\noindent In the Fig. \ref{Fig5} (a) and (b) we show $n^{ini}_a/N^2$ and 
$\langle \sigma^{ini}_a\rangle$ respectively as a function of $\dot{\gamma}$ for various system sizes. The system size effects are visible at lower shear rate limits but are within the sample to sample error bars. Using these values in equation 6, with $g=0.57$, we estimate the residual stress as a function of the shear-rate of the initial flow; see Fig. \ref{Fig6} (using blue symbols). Clearly, the residual stress predicted solely from the decay of active site at the flow cessation does not match with the numerically measured quantity. As discussed before, we observe production of new plastic events; hence, one has to take into account the occurrence and decay of the new active sites as well. We can quite easily extend the argument presented in the form of equation 4, except that for these newly activated sites we can drop the integral which took care fraction of relaxed active sites. The area under the curves presented in Fig.\ref{Fig4}(a) would give the total fraction newly activated sites $N^{new}_a/N^2$, which is shown in Fig. \ref{Fig5} (c). The associated stress per active sites $\langle \sigma^{rel}_a \rangle$ (averaged over the whole relaxation time) is shown in Fig. \ref{Fig5} (d). Hence the total stress associated with the newly created active sites is given by
\begin{equation}
    \sigma^{new}_{tot} = \frac{N^{new}_a}{N^2}\left \langle \sigma^{new}_a\right\rangle \frac{\tau \tau_{res}}{g\tau+\tau_{res}}
\end{equation}
\noindent Taking $\tau=1$ and $\tau_{res}=1$, the residual stress, including the contribution from the initially active sites and newly activated sites, can be written as
\begin{equation}
    \sigma_{R} = \sigma_{I} - \frac{n^{ini}_a}{N^2}\left\langle \sigma^{ini}_a \right\rangle \left( \frac{g - \log(1+g)}{g^2} \right ) - \frac{N^{new}_a}{N^2}\left\langle \sigma^{new}_a\right\rangle\left( \frac{1}{1+g} \right)
\end{equation}
\noindent The residual stress computed from the above equation is shown in the Fig. \ref{Fig6} (dark green triangles). We find that our simple semi-analytical estimation of residual stresses match quite well with the mesoscale simulation results.

\section{Characterising residual stress states}

After analysing the approach towards residual stress states, we now probe the mechanical properties of these states. Understanding the rheological properties of these states with frozen-in stress is quite relevant to applications and designing materials \cite{}. 

\subsection{Response to re-shearing}

\begin{figure}[t!]
\includegraphics[width=0.4500\textwidth]{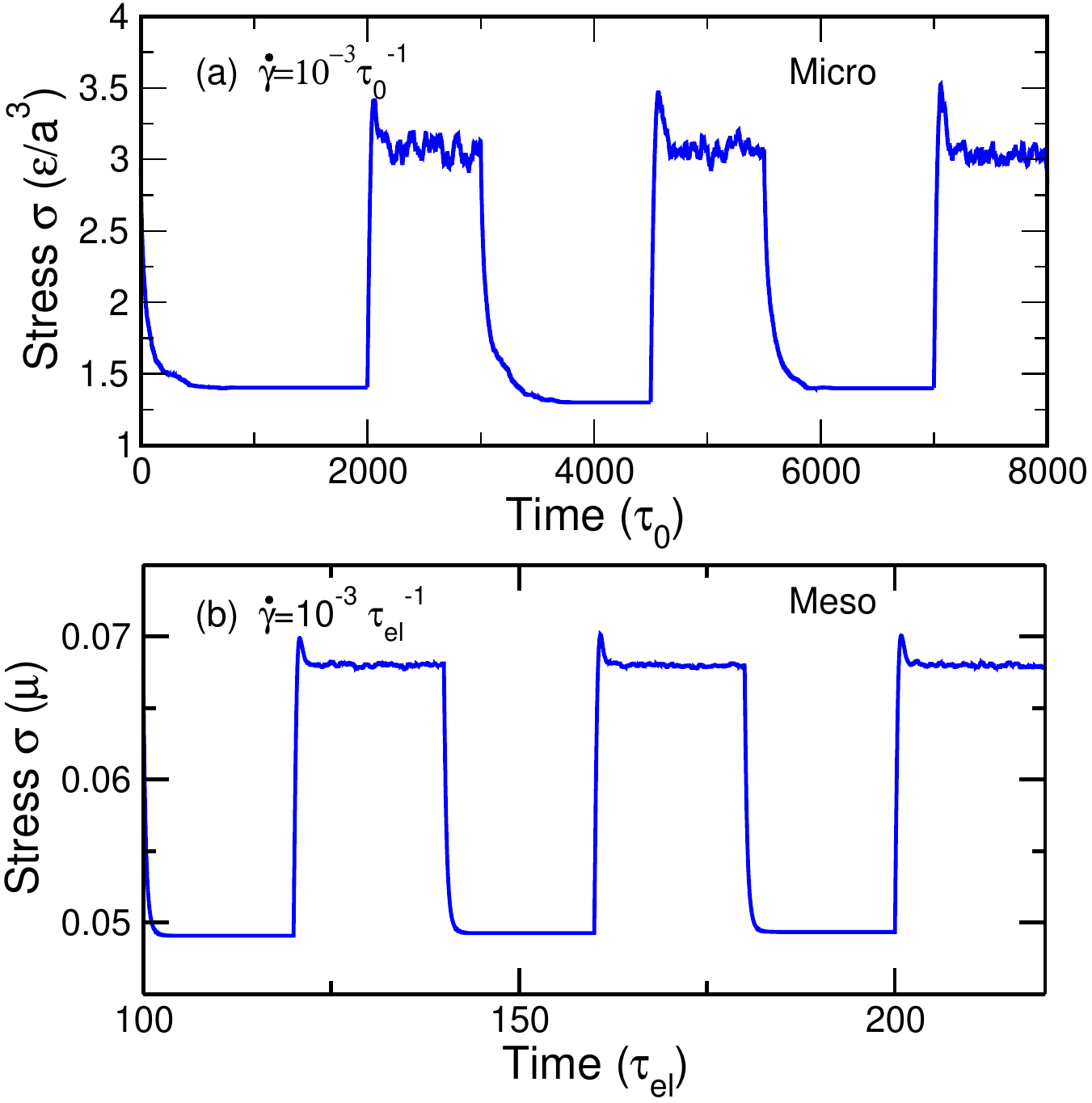}
\caption{Reproducible residual stress states, is demonstrated via the variation of stress with time, during a sequence of shear switch-off and switch-on using the same imposed shear-rate. Data in the top panel corresponds to microscopic model and data in the bottom panel corresponds to mesoscale model.}
\label{Fig7}
\end{figure}

We probe the response of the residual stress state by imposing a sequence of shear switch-on and switch-off, which we term as shear $\&$ relax cycles. The observed evolution of shear stress  due to this protocol is shown in Fig. \ref{Fig7} (a) and (b), for microscopic and mesoscale simulations respectively. 

Firstly, we note that if we impose the shear on the residual stress state using the same shear-rate as was done prior to the first switch-off and then again switch-off the applied shear, and continue with such shear $\&$ relax cycles, the intervening residual stress states obtained are always at the same level; see Fig. \ref{Fig7} (a) and (b). Some minor variations, observed in the microscopic simulations, are due to the fluctuations in the sampled states from the steadily flowing states. 

Secondly, we observe consistent stress overshoots during the shear switch-on part of the shear $\&$ relax cycles, in both microscopic as well as mesoscale simulations. This observation suggest that there is an unique transient shear response characteristic to a residual stress state. 
We note that the magnitude of the stress overshoot can be different from the one obtained from the shear of a amorphous state prepared from a thermal quench, where the history of the quench is relevant. Here, as we discuss later, it depends upon the residual stress level.

Overall, the above discussion highlights that residual stress states can be an ideal avenue for preparing amorphous states with reproducible response. Such states can be used to study in greater details the start-up response of disordered systems, both via elastoplastic as well as microscopic simulations.

\begin{figure}[b!]
\includegraphics[width=0.4500\textwidth]{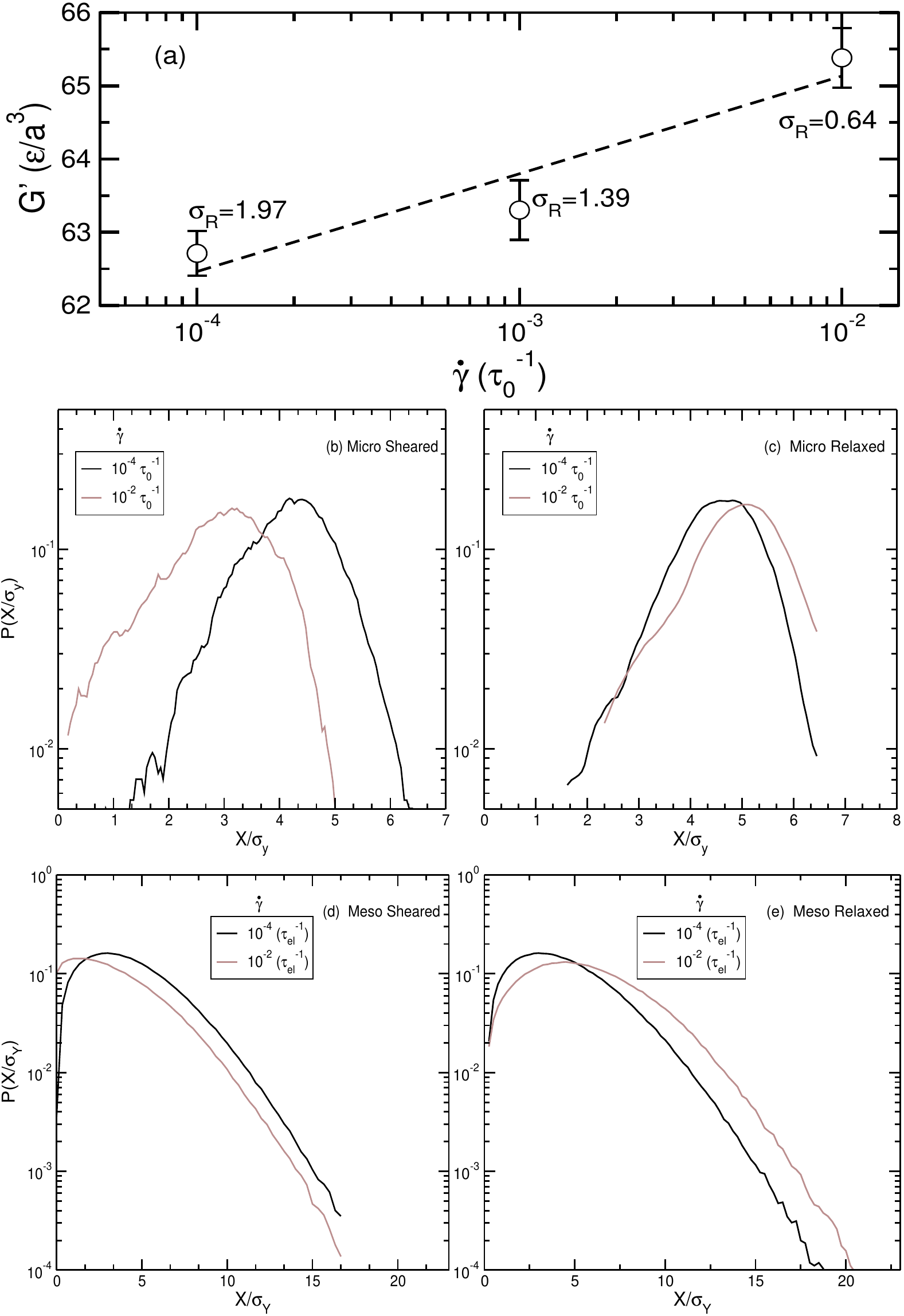}
\caption{Rigidity of residual of stress states. (Top) Variation of storage modulus, $G'$ of the arrested states, with the imposed shear-rate of the flowing state prior to switch-off, measured from the microscopic simulations. Also marked are the values of the obtained residual stresses, $\sigma_R$. (Bottom) Comparing local yield stress distribution for sheared state and corresponding residual stress state, for two different shear rates as marked. (a), (b) correspond to data from microscopic simulations
and (c), (d) correspond to data from mesoscale simulations.
}
\label{Fig8}
\end{figure}

\begin{figure}[th!]
\includegraphics[width=0.500\textwidth]{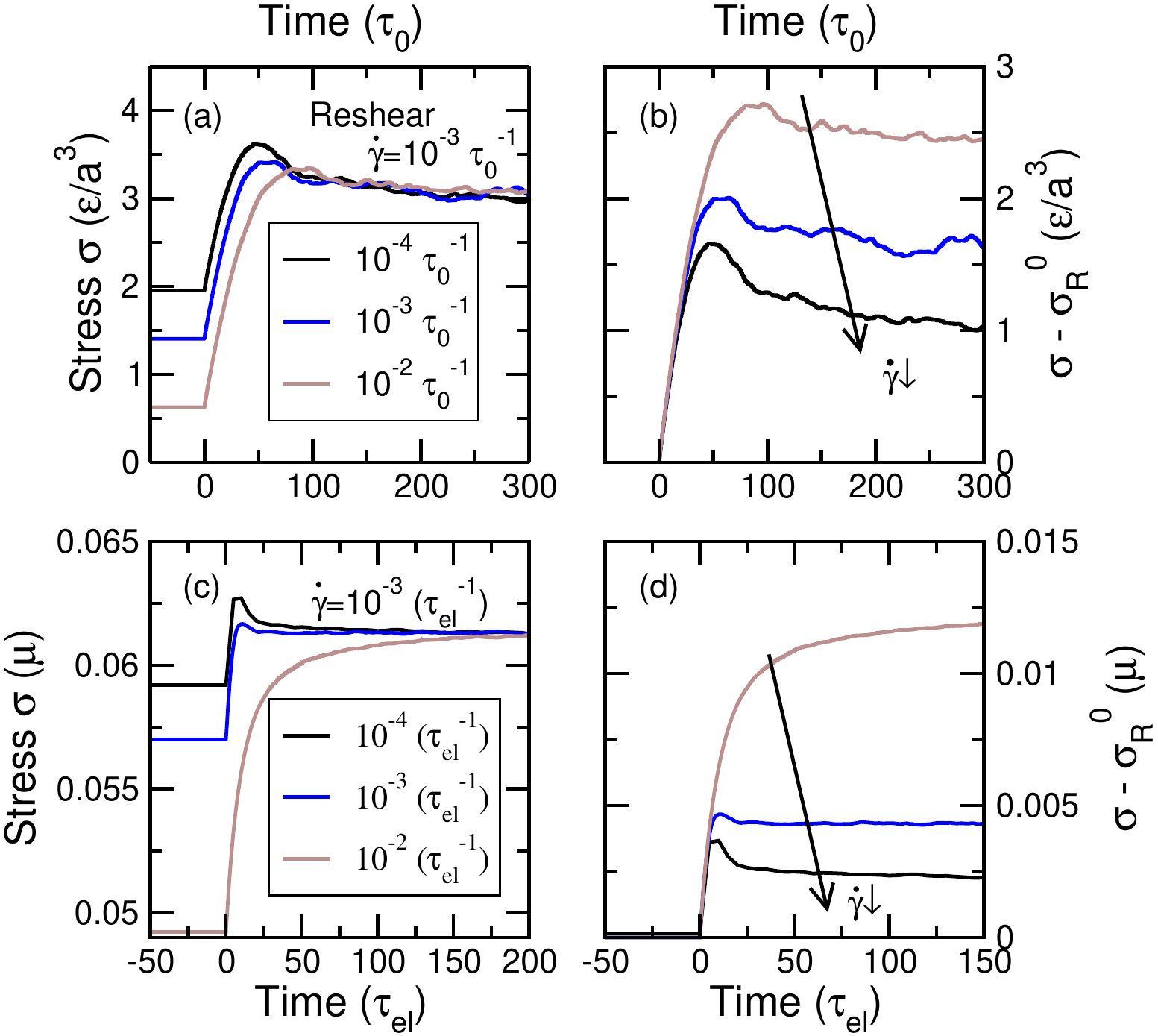}
\caption{(a) Response (stress vs strain curves) to imposed shear-rate ($\dot{\gamma}=10^{-3}\tau_0^{-1}1$) for states having different residual stress value, as indicated by the arrow. (b) The stress-strain curves, with the initial residual stress value deducted by $\sigma_R^0$, the residual stress value associated with $\dot{\gamma}=10^{-3}$ demonstrating that states having least residual stress demonstrates the largest stress overshoot.}
\label{Fig9}
\end{figure}

\subsection{Probing rigidity}

Next, we analyse the rigidity aspects of the residual stress states by computing the complex shear modulus using small amplitude oscillatory shear protocol. The storage modulus $G'$ obtained at low frequency limits are show in the Fig. \ref{Fig8} (a) as a function of $\dot{\gamma}$, the shear-rate of the flowing state prior to shear switch-off. We find that $G'$ decreases with increase in residual stress $\sigma_R$ (marked alongside the data points) or decrease in $\sigma_I - \sigma_R$. The loss modulus $G''$, which is around two orders of magnitude smaller, also show a similar dependence on the $\sigma_R$. For comparison, we also note that the $G'$ of the initial configuration, prepared via the thermal quenching at a particular cooling rate, is around $75 \epsilon/a^3$, which is higher than that of the residual stress state having the lowest $\sigma_R$ in our study. 

We further analyse the rigidity by computing the distribution of local yield thresholds $(X=\sigma_{lcl}^y -\sigma_{lcl}^0)$. In the microscopic simulations we use the frozen matrix method \cite{} to compute X. The distribution $P(X)$ computed (in both microscopic and mesoscale simulations) at the pre-sheared state as well as at residual state is shown in Fig.\ref{Fig8} (b-e) for two different $\dot{\gamma}$. Note that we have scaled the local yield thresholds by the macroscopic dynamic yield stress $\sigma_y$. In both microscopic and mesoscale models, we observe that the mean of P(X) is lower at higher $\dot{\gamma}$ of pre-sheared samples. This suggests that the residual states have higher shear rigidity than the flowing state, which is expected. The P(X) of residual states shows that mean is higher at higher pre-sheared $\dot{\gamma}$, suggesting that lower $\sigma_R$ generate higher mean yield threshold. This is a microscopic backing of what is observed in macroscopic storage modulus $G'$. Further analysis of local yield stress distribution both in understanding the local yielding phenomenon \cite{} as well as in the context of improving the mesoscale models \cite{} has to carried out systematically.

Having demonstrated that the low lying residual stress states have more rigidity, we now rationalise the transient response to the re-shear for a range of residual states. In Fig. \ref{Fig9} (a) and (c) we show three different residual states (obtained from three different pre-shear rates) subjected to re-shear at the same imposed shear rate. The three residual states show different transient response but eventually end up in the same steady state stress value. The stress overshoot obtained during the transient response show an initial residual stress state dependence. From the Fig. \ref{Fig9} (a) and (c) it seems like lower residual stress states show lower overshoot and is counter intuitive since the lower $\sigma_R$ corresponds to higher $G'$ and one expects from the previous studies \cite{} a higher stress overshoot. Once we subtract the initial residual stress ($\sigma - \sigma_R$) such that all the curves are compared at the same initial stress values (see Fig. \ref{Fig9} (b) and (c)) we find that the overshoot stress values increases with increase in rigidity state (or decrease in $\sigma_R$).

\section{Conclusion}
\label{section-conclusion}

In this work, we have investigated the origin of residual stresses in athermally driven amorphous solids after shear cessation, using a combination of microscopic and mesoscale elastoplastic models. 

We first demonstrated that the mesoscopic model qualitatively reproduces the phenomenology observed in microscopic simulations. In both cases,
we obtain arrested states exhibiting residual stresses after the forced shear flow is switched off. And, the monotonic dependence of the residual stress
on the shear stress of the flowing state under the applied shear, is also consistently observed. Thus, this provides the groundwork for using the mesoscale model to gain further insight into the formation of residual stresses, since these models have local plasticity as an intrinsic variable which allows for the identification as well as tracking of active sites, be it during shear or in the absence of it.

Thereafter, using the information provided by the mesoscale model, we develop a semi-analytical argument to provide a reasonable estimate of the eventually measured residual stress. The naive expectation is that when the applied shear is switched off, the active sites available in the system at that instant would release their stress and thereby the system  would reach the arrested state. We demonstrate, via the semi-analytical calculations, that such a scenario does not explain the level of residual stresses observed in the simulations. On the other hand, the spatial mesoscale maps,
constructed from the cumulative stress relaxation following the shear switch-off, reveal that stress relaxation not only happens at these active sites but also in the surroundings even long after the forcing is switched off. This implies that new plastic events occur during the stress relaxation, which contribute to the observation of new active sites. The number of these newly activates sites eventually relaxes after a given time, and thereby the system reaches the final state with a finite residual stress. The scale of the cascade of the new events depends upon the initial stress in the steady flow, the larger the initial stress, the more activity is generated during the relaxation following shear switch off which leads to a lower residual stress. When the occurrence of the newly activated events are appropriately included within the semi-analytical calculations, we can reasonably reproduce the scale of residual stresses observed in the simulations. 

Finally, we have studied the mechanical response of the residual stress states. We first observe, by imposing a sequence of steady deformation and relaxation cycles, that the level of residual stress is reproducible, expectedly, for a state flowing under a particular shear-rate prior to shear switch-off. More importantly this sequence also enabled us to demonstrate that the transient shear response of these residual stress states is also reproducible, with the same level of stress overshoots in each cycle. This is very significant for preparation of amorphous states with reproducible rheological response. We also probed the rigidity of the residual stress states in two ways, viz. by measuring the storage modulus via oscillatory shear, and then by measuring local yield stress distributions. Both protocols revealed that the states having lower residual stresses are more rigid and have a higher yield threshold. Consequently, when we impose the same shear-rate to states having different residual stresses, the state having the least residual stress exhibits the largest stress overshoot. All these observations are consistent across microscopic and mesoscale models.

\section{Acknowledgements}

All authors acknowledge financial support from CEFIPRA Grant No.~5604-1 (AMORPHOUS-MULTISCALE).
K.~M. acknowledges financial support of the French Agence Nationale de la Recherche (ANR), under grant ANR-14-CE32-0005 (FAPRES).
We thank Jean-Louis Barrat, Emanuela Del Gado and Alberto Rosso for useful discussions.

\bibliographystyle{rsc} 

\bibliography{residual-stress}



\end{document}